\documentclass[12pt,letter]{article}
\pdfoutput=1
\usepackage{graphicx, epsfig, color,cite,inputenc}
\usepackage{amsmath}
\usepackage{amssymb}
\usepackage{float}
\usepackage{caption,subcaption,graphicx}
\usepackage{hyperref}

\def\to{\rightarrow}

\def\bi{\begin{itemize}}
\def\ei{\end{itemize}}

\def\alt{\lesssim}
\def\agt{\gtrsim}
\def\be{\begin{equation}}  
\def\ee{\end{equation}}  
\def\bea{\begin{eqnarray}}  
\def\eea{\end{eqnarray}}

\begin{document}
\begin{titlepage}
\begin{flushright}
OU-HEP-211110
\end{flushright}

\vspace{0.5cm}
\begin{center}
{\Large \bf An anthropic solution\\ to the cosmological moduli problem
}\\ 
\vspace{1.2cm} \renewcommand{\thefootnote}{\fnsymbol{footnote}}
{\large Howard Baer$^{1,2}$\footnote[1]{Email: baer@ou.edu },
Vernon Barger$^2$\footnote[2]{Email: barger@pheno.wisc.edu} and
Robert Wiley Deal$^{1,2}$\footnote[3]{Email: rwileydeal@ou.edu} 
}\\ 
\vspace{1.2cm} \renewcommand{\thefootnote}{\arabic{footnote}}
{\it 
$^1$Homer L. Dodge Department of Physics and Astronomy,
University of Oklahoma, Norman, OK 73019, USA \\[3pt]
}
{\it 
$^2$Department of Physics,
University of Wisconsin, Madison, WI 53706 USA \\[3pt]
}

\end{center}

\vspace{0.5cm}
\begin{abstract}
\noindent
Light moduli fields, gravitationally coupled scalar fields with no classical potential 
and which are expected to emerge as remnants from string theory compactification,
are dangerous to cosmology in that 
1. their late-time decays may disrupt successful Big Bang Nucleosynthesis (BBN), 
2. they may decay into gravitino pairs which result in violation of BBN constraints 
or overproduction of lightest SUSY particles (LSPs, assumed to constitute at least a portion of the dark matter in the universe) 
and 
3. they may decay directly into LSPs, resulting in gross DM overproduction.
Together, these constitute the cosmological moduli problem (CMP).
The combined effects require lightest modulus mass $m_\phi \agt 10^4$ TeV, and if the 
lightest modulus mass $m_\phi$ is correlated with the SUSY breaking scale $m_{3/2}$, 
then the underlying SUSY model would be highly unnatural. 
We present a solution to the CMP wherein the lightest modulus initial field
strength $\phi_0$ is anthropically selected to be $\phi_0\sim 10^{-7}m_P$ by the 
requirement that the dark matter-to-baryonic matter ratio be not-too-far removed
from its present value so that sufficient baryons are present in the universe 
to create observers. In this case, instead of dark matter overproduction via 
neutralino reannihilation at the modulus decay temperature, 
the neutralinos inherit the reduced moduli number density, 
thereby gaining accord with the measured dark matter relic density.
\end{abstract}
\end{titlepage}


Moduli fields are gravitationally coupled scalar fields with no classical 
potential which arise from string theory compactifications from 10 to 4
spacetime dimensions\cite{Ibanez:2012zz}. 
Upon stabilization of the moduli, their vevs determine
many aspects of the resulting 4-d low energy effective field theory (EFT)
such as gauge and Yukawa couplings and the scale of supersymmetry breaking.
In II-B string theory compactifications, one expects Hodge number $h^{2,1}$
complex structure moduli $U^\beta$ and $h^{1,1}$ K\"ahler moduli $T^\alpha$ which parametrize metric deformations under Calabi-Yau compactifications. These are
accompanied by the non-geometric axio-dilaton modulus $S$. 
Under typical compactifications, the Hodge numbers are expected to be of order
several hundred: $h^{2,1}\sim h^{1,1}\sim {\cal O}(100)$\cite{Kreuzer:2000xy}.

Under flux compactifications\cite{Douglas:2006es}, 
the $U^\beta$ and $S$ fields can be stabilized by flux
to gain Kaluza-Klein (KK) scale masses while the K\"ahler moduli
are expected to be stabilized by non-perturbative effects (KKLT)\cite{Kachru:2003aw} 
or a combination of perturbative/non-perturbative effects under the 
large volume scenario (LVS)\cite{Balasubramanian:2005zx}. 
The K\"ahler moduli may have masses extending as low as the weak scale 
and arguments have been made that the 
lightest of the K\"ahler moduli $\phi$ should have mass comparable to the 
gravitino mass $m_\phi\sim m_{3/2}$\cite{Denef:2004cf,Acharya:2010af}. 
Such light moduli fields present a  significant danger to the standard cosmology, a situation known as the cosmological moduli problem (CMP)\cite{Kane:2015jia}.

We consider the case of scalar field evolution just after the end of inflation
when the universe has (re)-heated to a temperature $T_R$.
Scalar fields $\phi$ in the early universe obey the wave equation
$\ddot{\phi}+3H\dot{\phi}+V^\prime =0$ where the dots denote time derivatives, 
$H(T)$ is the Hubble parameter, $V$ is the scalar potential, and the prime denotes
derivative with respect to $\phi$. In the very early universe with large $H$, 
the friction term $3H\dot{\phi}$ is dominant and the value of $\phi$ is
effectively frozen. As the universe expands and $H$ decreases, the potential
can be approximated as a simple harmonic oscillator and the $\phi$ field
begins to oscillate at a temperature $T_{osc}\simeq (10/\pi^2g_*(T_{osc}))^{1/4}\sqrt{m_Pm_{\phi}}$ where $m_P$ is the reduced Planck mass
and $g_*$ is the effective degrees of freedom parameter at temperature $T$.
The oscillating $\phi$ field behaves as cold dark matter and dilutes under expansion
more slowly than radiation and hence comes to dominate the energy density of the universe
at the radiation equality temperature $T_e=(3/2m_P^2)\phi_0^2\sqrt{m_Pm_\phi}(10/\pi^2g_*(T_e))^{1/4}$ (where $\phi_0$ is the magnitude of the lightest modulus field
which is set during inflation). Eventually, the $\phi$ field decays at a radiation 
temperature $T_D=\sqrt{m_P\Gamma_\phi}/(\pi^2g_*(T_D)/90)^{1/4}$ into all available modes.

\begin{figure}[tbh]
\begin{center}
\includegraphics[height=0.5\textheight]{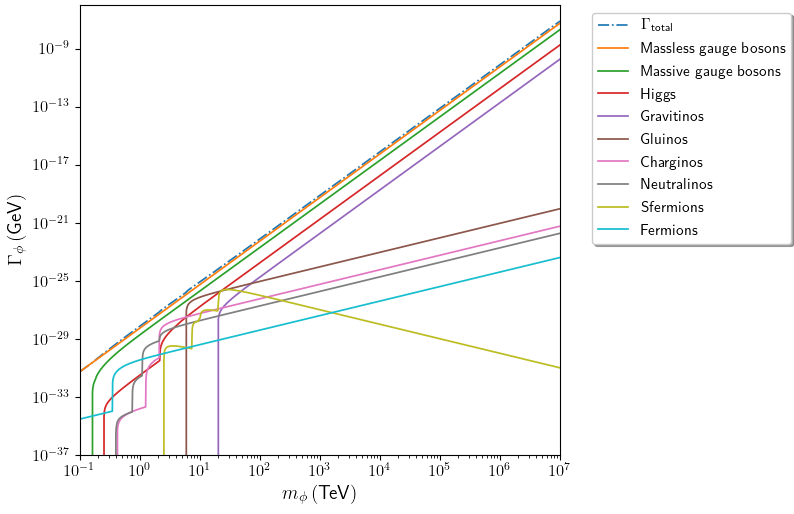}
\caption{Lightest modulus field partial decay widths into MSSM particles and gravitinos
vs. $m_{\phi}$ (case with suppressed decay to gauginos but unsuppressed decays to gravitinos with all MR couplings $\lambda_i =1$).
\label{fig:Gamma_phi}}
\end{center}
\end{figure}

Here, given the substantial motivations for weak scale supersymmetry, 
especially with regard to stabilizing radiative corrections to scalar field masses
such as the Higgs boson, we will assume a low energy EFT as described by the minimal
supersymmetric standard model or MSSM\cite{Baer:2006rs}. 
In this case, the lightest modulus $\phi$ decays
to SM particles (radiation), SUSY particles, and gravitinos. We have computed all
2-body MSSM decay modes of the $\phi$ field using the Moroi-Randall\cite{Moroi:1999zb} 
operators which are proportional to coupling parameters $\lambda_i$, where $i$ labels
the various modulus couplings to gauge bosons/gauginos, Higgs superfields, and matter
superfields. The results will be reported in a longer paper\cite{bbbw} but 
example results are shown in Fig. \ref{fig:Gamma_phi} which shows the
modulus field partial widths as a function of $m_{\phi}$ for all couplings 
$\lambda_i=1$ in case of helicity-suppressed decays to gauginos but unsuppressed
decays to gravitinos\cite{Asaka:2006bv,Nakamura:2006uc,Dine:2006ii}. 
This plot is shown assuming a natural SUSY benchmark point\cite{bbbw}.
Once $\Gamma_\phi$ is known, the modulus decay temperature $T_D$ 
and various branching fractions can be computed.

\begin{figure}[tbh]
\begin{center}
\includegraphics[height=0.5\textheight]{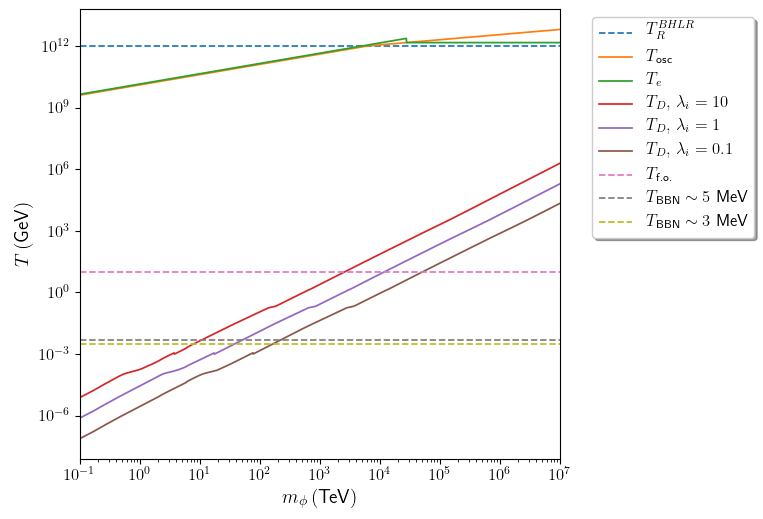}
\caption{Modulus decay temperatures $T_D$ vs. $m_{\phi}$ for three values of the MR 
couplings $\lambda_i$ along with LSP freeze-out temperature $T_{fo}$ and $T_{BBN}$.
We also show a putative value for $T_R\sim 10^{12}$ GeV along with modulus
oscillation temperature $T_{osc}$ and modulus field-radiation equality 
temeprature $T_e$.
\label{fig:TD}}
\end{center}
\end{figure}

Originally, the CMP was concerned with late decays of scalar fields in the 
early universe wherein the decay energy release, if occurring after the onset
of Big Bang Nucleosynthesis (BBN), would disrupt the production of light 
elements\cite{Coughlan:1983ci,Banks:1993en,deCarlos:1993wie}.
As a rough estimate, we take the temperature $T_{BBN}\sim 3-5$ MeV at which the onset of 
BBN starts.\footnote{More precise estimates range from 1.8-5 MeV, 
see {\it e.g.} Refs. \cite{Kawasaki:1999na,Kawasaki:2000en,Hasegawa:2019jsa}.} 
The $T_{BBN}$ is shown by the two lower horizontal lines in
Fig. \ref{fig:TD} along with the modulus field decay temperature $T_D$  for
three values of MR couplings $\lambda_i=0.1$, 1, and 10. 
From the plot, we see that the modulus decay temperature $T_D$ exceeds $T_{BBN}$
for modulus mass values $m_{\phi}\gtrsim 10-100$ TeV. Lighter values of 
$m_\phi$ are then excluded by BBN constraints. For perspective, we also show at the top
of the plot the values of $T_{osc}$ and $T_e$. Since $T_e$ depends on the initial
modulus field strength $\phi_0$, which is expected to be set by inflationary 
cosmology to values of order $\phi_0\sim m_P$, its values are very high, in 
$10^9-10^{12}$ GeV range. For oscillations beginning after the inflationary reheating period, 
$T_{osc} \sim \sqrt{m_P m_\phi}$ and hence typically gives values of the same order of magnitude as $T_e$.
We show also a putative re-heat temperature $T_R\sim 10^{12}$
GeV which is right at the Buchmuller limit $T_{BHLR}$\cite{Buchmuller:2004xr}, above which thermal 
effects can destabilize the dilaton effective potential and give rise to a runaway
dilaton field $S$.

However, even if $T_D>T_{BBN}$, we have still not necessarily solved the CMP 
problem. A further problem arises if the $\phi$ field decays at a substantial
rate into gravitinos $\psi_\mu$\cite{Kawasaki:2004qu,Hashimoto:1998mu,Asaka:2006bv,Nakamura:2006uc,Endo:2006zj,Dine:2006ii,Kawasaki:2017bqm}. 
These decays may or may not be helicity 
suppressed depending on details of the supergravity K\"ahler potential\cite{Dine:2006ii}. 
Then the gravitinos also decay late and can disrupt BBN,
in which case we need knowledge of $T_{3/2}$, the gravitino decay temperature.
Furthermore, the gravitinos can decay to lightest SUSY particles (LSPs, denoted as $\chi$)
and potentially overproduce dark matter (assumed to be the $\chi$ particles).
Further details will be reported in Ref. \cite{bbbw}, but for the moment we will
assume $m_\phi\sim m_{3/2}$ so that the decay mode $\phi\to\psi_\mu\psi_\mu$ is
kinematically closed as suggested by Acharya {\it et al.} Ref. \cite{Acharya:2010af}.

A further concern for the CMP is the direct overproduction of dark matter $\chi$
from moduli cascade decays. The initial modulus energy density is 
$\rho_\phi=m_\phi^2\phi_0^2/2$ and the modulus number density is roughly 
$n_\phi\sim \rho_\phi/m_\phi$. Assuming each modulus particle decays 
with branching fraction $B(\phi\to\chi)$, and accounting for the expansion
of the universe between $T_{osc}$ and $T_D$, we find
\be
n_\chi^D\sim B(\phi\to\chi )m_\phi \phi_0^2\left(\frac{g_*(T_D)T_D^3}{g_*(T_{osc})T_{osc}^3}\right),
\label{eq:nD}
\ee 
{\it i.e.} naively, the neutralinos initially inherit the modulus field number density, 
subject to branching ratio and expansion effects.
Of course, if the modulus decay temperature $T_D$ exceeds the LSP
freeze-out temperature $T_{fo}\sim m_\chi/20$, then the neutralinos will
thermalize and their relic density will be just the standard 
thermally-produced (TP) result $\Omega_\chi^{TP}h^2$.

However, this accounting can be greatly modified if the modulus field 
decays at temperature $T_D<T_{fo}$ and the number density 
$n_\chi$ exceeds the critical density $n_\chi^c$ above which 
neutralino {\it reannihilation effects} may be important\cite{Giudice:2000ex,Choi:2008zq,Baer:2011hx}. The Boltzmann
equation for the neutralino number density is given by
\be
\frac{dn_\chi}{dt}+3Hn_\chi=-\langle\sigma v\rangle n_{\chi}^2
\ee
and if $\langle \sigma v\rangle n_\chi(T_D)>H(T_D)$, then neutralinos
will reannihilate after modulus decay. The Boltzmann equation, rewritten
in terms of the yield variable $Y_\chi\equiv n_\chi/s$, where $s$ is the
entropy density ($s=\frac{2\pi^2}{45}g_{*S}T^3$ for radiation) is given by
\be
\frac{dY_\chi}{dt}=-\langle\sigma v\rangle Y_\chi^2 s
\ee
which, assuming $\langle\sigma v\rangle$ is dominated by the constant term, 
can be easily integrated to find $Y_\chi^{reann}\simeq H(T_D)/\langle\sigma v\rangle s(T_D)$ or
\be
n_\chi^c \simeq H/\langle\sigma v\rangle |_{T=T_D}
\ee
so that 
\be
\Omega_\chi^{reann}h^2\simeq \Omega_\chi^{TP}h^2(T_{fo}/T_D),
\ee
{\it i.e.} the reannihilation abundance is enhanced from its TP value
by a factor $T_{fo}/T_D$. The final neutralino number density is then given by
\be
n_\chi\sim \min\left\{ n_\chi^c,n_\chi^D\right\}
\label{eq:nc}
\ee
with $\Omega_\chi h^2= m_\chi n_\chi/\rho_ch^{-2}$
where $\rho_c$ is the critical closure density and $h$ is the scaled Hubble constant.
Provided reannihilation effects are important, the reannihilation
enhancement factor $T_{fo}/T_D$ can be read off from Fig. \ref{fig:TD}.
Since $\phi_0$ is so large, almost always $n_\chi^c<n_\chi^D$ so that the
neutralino relic abundance takes the reannihilation value for $T_D<T_{fo}$.

\begin{figure}[tbh]
\begin{center}
\includegraphics[height=0.5\textheight]{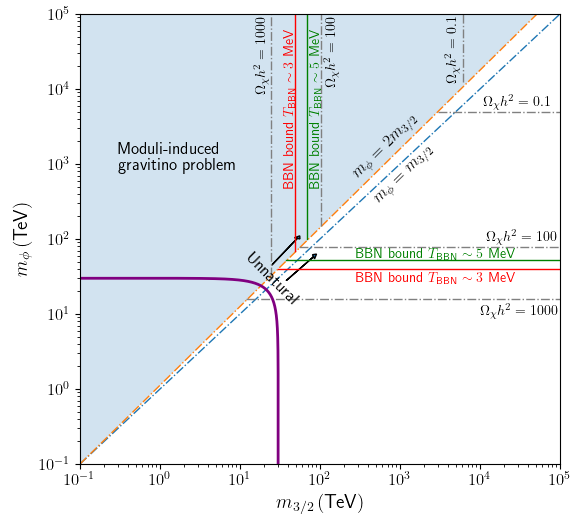}
\caption{Regions of the $m_{3/2}$ vs. $m_\phi$ plane which are allowed 
(upper-right with $\Omega_{\chi}h^2<0.1$) or excluded. 
We also show BBN bounds and an approximate curve (purple) beneath which 
SUSY may be natural. The shaded region is subject to the moduli-induced gravitino problem.
\label{fig:plane}}
\end{center}
\end{figure}

The overall picture is given in Fig. \ref{fig:plane}, where we plot the allowed and
excluded regions in the $m_{3/2}$ vs. $m_\phi$ plane. 
The upper-left shaded region is plagued by the moduli-induced gravitino problem.
We also see that one must live to the upper-right of the BBN bounds
which set in around $m_{3/2},\ m_\phi\sim 30$ TeV. However, even if one avoids the
BBN bounds by taking heavy gravitino/moduli masses, the overproduction of dark matter is
considerably above measured values unless $m_\phi\agt 5\times 10^3$ TeV.
For comparison, we include a purple circle of naturalness favored region\cite{Baer:2012cf}
where $\sqrt{m_\phi^2+m_{3/2}^2} <30$ TeV. The CMP-consistent regions are
far beyond the naturalness regions of parameter space. 

A variety of solutions to the CMP have been proposed. 
Some intriguing possibilities include 
1. a period of late weak scale\cite{Randall:1994fr} or thermal inflation\cite{Lyth:1995ka} which
could dilute all relics after moduli decays are complete, 
2. the presence of a late decaying saxion field which could dilute all relics\cite{Endo:2006ix} 
(but one must then beware of dark matter/BBN constraints from the associated 
axino field\cite{Baer:2011hx}),
3. additional hidden sector fields connected to the visible sector via a portal
coupling so that the dark matter is actually a hidden sector semi-inert 
particle\cite{Blinov:2014nla,Acharya:2016fge}
and 4. abandoning naturalness and allowing for a thermally underproduced 
wino LSP where then the modulus decay into winos makes up the 
dark matter deficit\cite{Moroi:1999zb,Acharya:2008bk}.
This latter scenario now seems excluded by rather strict indirect detection limits
on wino dark matter\cite{Cohen:2013ama,Fan:2013faa,Baer:2016ucr}.

Here, we propose an alternative anthropic solution to the CMP. It is somewhat
similar to an older proposal for an anthropic solution to 
overproduction of axion dark matter in stringy scenarios with a large axion
decay constant $f_a\sim 10^{16}$ GeV\cite{Linde:1987bx,Wilczek:2004cr,Tegmark:2005dy,Freivogel:2008qc}. 
In those papers, it is considered
that there is an anthropic upper bound on the ratio $r\equiv \rho_{DM}/\rho_{baryons}$
where if the dark matter energy density $\rho_{DM}$ is too high,
then the dominant structures that form in the early universe will be virialized clouds
of dark matter and baryons are present at such a low rate that galaxies and hence
stars and observers as we know them will not form. 
The exact upper bound on $r$ is difficult
to calculate but attempts have been made in Aguire {\it et al.}\cite{Tegmark:2005dy}
and by Freivogel\cite{Freivogel:2008qc} using Bousso's causal diamond measure\cite{Bousso:2006ev} and by Bousso and Hall\cite{Bousso:2013rda}. 
Typically, it is expected that
the dark matter-to-baryon ratio $r$ should be $\sim 1-10$ (whereas observation
sets it at $\sim 5$). For a uniform distribution of axion misalignment angles
$\theta$ distributed across the decades of possibilities, then anthropics may 
require small $\theta$ leading to $\Omega_{axion}h^2\sim 0.1$ even for
string scale axion decay constants.

In our approach, we note that for much of the plane shown in Fig. \ref{fig:plane}, 
dark matter in the form of WIMPs is grossly overproduced by several orders of magnitude.
This is especially true in the lower-left natural region  of $m_\phi$ vs. $m_{3/2}$
space. Here, we expect that the landscape also sets the magnitude of the 
weak scale in various pocket-universes to be $\sim 100$ GeV. This occurs from
bounds derived by Agrawal {\it et al.}\cite{Agrawal:1998xa} that if the pocket-universe value of the 
weak scale is displaced by more than a factor $2-5$ from its measured value, 
then the proton-neutron mass difference is displaced in such a fashion as not to
allow complex nuclei, and hence atoms as we know them, 
to form (the atomic principle\cite{Arkani-Hamed:2005zuc}).
Natural SUSY is expected to be far more prevalent on the landscape than unnatural
models because in unnatural models, one would have to land on finely tuned
parameter choices to keep the weak scale weak, whereas in natural models a wide
range of natural parameters are allowable\cite{Baer:2019cae}. 
It is expected that soft SUSY breaking
parameters obey either a power-law\cite{Susskind:2004uv,Douglas:2004qg,Arkani-Hamed:2005zuc} or log\cite{Broeckel:2020fdz} selection favoring large
values subject to the anthropic condition of the atomic principle. Scans over
the landscape\cite{Baer:2017uvn} then find the Higgs mass $m_h$ pulled up to a statistical peak
around $m_h\sim 125$ GeV (due in part to large mixing in the stop sector)
while sparticles are pulled beyond present LHC search limits. 
The first/second generation matter scalars are pulled into the $10-40$ TeV range
since they have only suppressed contributions to the weak scale. The maximal scalar
masses should be comparable to the gravitino mass $m_{3/2}$ in gravity-mediated SUSY breaking models. Thus, we expect  a corresponding value $m_{3/2}\sim 10-40$ TeV
in natural SUSY from the landscape.

\begin{figure}[tbh]
\begin{center}
\includegraphics[height=0.5\textheight]{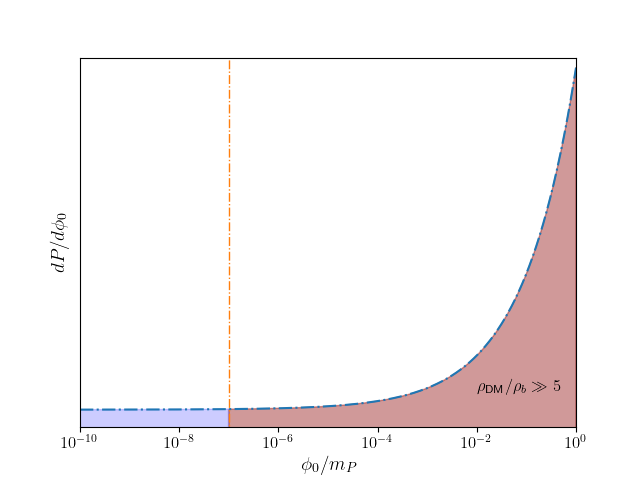}
\caption{A putative distribution of $\phi_0$ values from the string landscape.
Values of $\phi_0$ giving rise to $\rho_{DM}/\rho_{baryons}\gg 5$ are anthropically
vetoed leading to a value of $\phi_0\sim 10^{-7}m_P$.
\label{fig:phi0}}
\end{center}
\end{figure}

Thus, in the case of landscape SUSY, which is typified by a gravitino mass
and soft breaking scale $m_{3/2}\sim 30$ TeV, we will also require no great 
overproduction of dark matter compared to baryons in anthropically-allowed 
pocket universes. We will also 
\bi
\item assume that the lightest moduli field strength
$\phi_0$ is spread somewhat uniformly in value across the decades, although the
volume of $\phi_0$ space favors large $\phi_0$ values (as in Weinberg's
solution to the cosmological constant problem)\cite{Weinberg:1987dv}. 
Thus, we feign ignorance about the particular inflationary model which gives rise to
$\phi_0$ so long as it includes a distribution of $\phi_0$ values extending well
below $m_P$ that is somewhat uniformly distributed in the area of selection.
\ei
Then, large values of $\phi_0\sim m_P$ are certainly included in the distribution
$P(\phi_0 )$ but will be {\it vetoed}  due to overproduction of dark matter 
which leads to unlivable pocket universes: 
one will have an anthropic selection of much smaller $\phi_0$ values.
The situation is illustrated in Fig. \ref{fig:phi0}, where we show a putative 
landscape distribution $dP/d\phi_0$ vs. $\phi_0$. Values of $\phi_0$ in the red region
lead to too large a value of $\rho_{DM}/\rho_{baryons}$ and hence to unlivable
pocket universes. 
The distribution favors pocket universes with $\rho_{DM}/\rho_{baryons}$ as large as
possible such that baryonic structures such as galaxies and stars can form, leading to livable pocket universes.

How might such a distribution as Fig. \ref{fig:phi0} arise in the 
eternally inflating multiverse? To see how, we adopt a toy example from 
Dine, Randall and Thomas (DRT) Ref. \cite{Dine:1995kz} where during inflation we expect the scalar potential for the lightest modulus $\phi$ to have the form
\be
V_{inf}(\phi)=(m_{soft}^2-a^2H_I^2)|\phi|^2+\frac{1}{2m_P^2}(m_{soft}^2+b^2H_I^2)|\phi|^4
\ee
where the $m_{soft}$ terms come from usual SUGRA breaking with $m_{soft}\sim m_{3/2}$ 
but the terms involving $H_I\sim \sqrt{V_{inf}}$ are {\it inflation-induced} soft term contributions to the scalar potential. 
They arise because the inflaton potential $V_{inf}(\phi_I)$ (where $\phi_I$ is the inflaton field) necessarily breaks SUSY (since $V_{inf}> 0$ during inflation)\cite{Dine:1995uk,Dine:1995kz}.
For $H_{inf}\gg m_{soft}$, then the minimum of the potential occurs for 
$\phi_{min1}\sim (a/b)m_P$  while for $H\ll m_{soft}$, then the minimum occurs at 
$\phi_{min2}\sim m_P$. The difference sets the value of $\phi_0$: 
$\phi_0\sim (1-(a/b))m_P$. Thus, the expectation for $\phi_0$ is at $\sim m_P$,
but with $\sim 10^{500}$ pocket universes\cite{Ashok:2003gk} and with $a$ and $b$ 
scanning over different values in the multiverse, there are still 
considerable numbers of pocket universes with $(a/b)\sim 1$ so that 
$\phi_0\ll m_P$. This is the picture shown in Fig. \ref{fig:phi0}. 
DRT actually suggested a solution to the CMP wherein the presence of some 
hidden symmetry causes the inflationary minimum to coincide with the post-inflationary minimum, thus setting $\phi_0$ to tiny values. 
In our case, there would be no symmetry but instead anthropic conditions veto
the solutions with $\phi_0$ too big. In this sense, the proposed DRT solution is akin to the conjecture of some symmetry which sets the cosmological constant to zero, whilst our solution is akin to Weinberg's wherein only tiny cosmological constants are allowed by
anthropic conditions.

For tiny $\phi_0$, the moduli number density is highly suppressed since it depends on 
$\phi_0^2$, and so few $\chi$ particles are produced
in $\phi$ decays that they fail to meet the reannihilation criterion. 
In this case, the final neutralino number density follows Eq. \ref{eq:nD}.

\begin{figure}[tbh]
\begin{center}
\includegraphics[height=0.5\textheight]{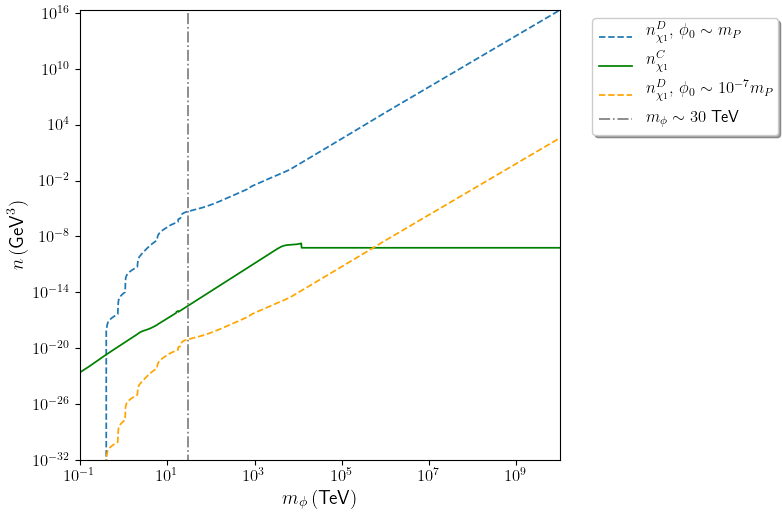}
\caption{LSP number density $n_{\chi}^D$ from modulus decay along with
critical number density $n_{\chi}^c$ which is expected to arise from
LSP reannihilation after modulus decay in the early universe.
\label{fig:nchi}}
\end{center}
\end{figure}

In Fig. \ref{fig:nchi}, we show the neutralino critical number density
$n_\chi^c$ as the green curve, along with the decay-produced
number density $n_\chi^D$ with $\phi_0\sim m_P$ (blue dashed curve). Since the $n_\chi^c$ 
curve is almost always below the blue-dashed $n_\chi^D$ curve, then reannihilation
rates produce a large overabundance of LSP dark matter. But if $\phi_0$ is
anthropically selected at $\sim 10^{-7}m_P$, then we obtain instead the
orange-dashed curve for $n_\chi^D$. It is well below the $n_\chi^c$ curve and so the
neutralinos inherit the moduli field number density (subject to the 
lightest modulus branching fraction to SUSY particles). This suppression
brings the neutralino relic density down to the $\Omega_\chi^{NTP}h^2\sim 0.1$
level. This solves the CMP.

Actually, in natural SUSY, which we assume here, the lightest neutralino is
expected to be dominantly higgsino-like, since the superpotential $\mu$ 
parameter feeds directly into the value of the weak scale, but also sets the mass
of the higgsinos\cite{Chan:1997bi,Baer:2011ec}. In Ref. \cite{Baer:2018rhs}, it is shown that higgsino-only dark matter
for $m_\chi\alt 350$ GeV (the naturalness region) is likely excluded by direct and 
indirect detection bounds. Likewise, in Ref. \cite{Han:2019vxi}, it is claimed
higgsino-only DM with mass $m_\chi\alt 550$ GeV is excluded by AMS searches.
A way out is that one must also require naturalness in the QCD sector, which
addresses the strong CP problem. Then the SUSY DM is expected to be
an axion-WIMP admixture\cite{Baer:2011hx}, where axions typically dominate the higgsino abundance.
In this case, DD and IDD dark matter limits are considerably weakened since
WIMPs only make up a small portion of dark matter\cite{Baer:2013vpa}. 
For our case here, if $\phi_0$ is
selected somewhat lower, then there is room left for axions and natural 
higgsino-like WIMPs remain as allowed DM candidates. For the case of SUSY axions, 
we needn't rely on anthropics since the soft SUSY breaking terms that enter the
PQ potential set the axion scale to $f_a\sim 10^{11}$ GeV\cite{Baer:2019uom}.
In future work, we will consider the more model-dependent effect of
dark radiation from moduli decay\cite{Cicoli:2012aq,Higaki:2012ar,Higaki:2013lra,Conlon:2013isa,Allahverdi:2014ppa,Cicoli:2015bpq,Takahashi:2019ypv,Reig:2021ipa}.

{\it Conclusion:} We have computed all lightest modulus decay modes including
mixing and phase space factors and used these results to examine the
cosmological moduli problem with respect to BBN and dark matter and gravitino
overproduction. In the case where gravitino production is not allowed
($m_\phi<2m_{3/2}$), then the most severe constraint comes from overproduction of WIMPs
via late time moduli decay. One resolution to the CMP requires a lightest modulus
mass $m_\phi\agt 5\times 10^3$ TeV. If the modulus mass comes from SUSY breaking and is
$\sim m_{3/2}$, then such a huge modulus mass brings extreme tension with naturalness.
We present an alternative solution to the CMP wherein initial modulus
field values are anthropically selected to suppressed values $\phi_0\alt 10^{-7}m_P$
in order to avoid overproduction of dark matter via modulus decay. The anthropic solution
allows for light moduli and gravitino masses to exist in the naturalness preferred regime of values $m_\phi\sim m_{3/2}\sim 30$ TeV.

{\it Acknowledgements:} 

This material is based upon work supported by the U.S. Department of Energy, 
Office of Science, Office of High Energy Physics under Award Number DE-SC-0009956 and U.S. Department of Energy (DoE) Grant DE-SC-0017647. 


\bibliography{mod0}
\bibliographystyle{elsarticle-num}

\end{document}